\documentclass[sigconf,nonacm,9pt]{acmart}
\AtBeginDocument{%
  }
\renewcommand\footnotetextcopyrightpermission[1]{} 
\setcopyright{acmlicensed}
\copyrightyear{2018}
\acmYear{2018}
\acmDOI{XXXXXXX.XXXXXXX}
\acmConference[Conference acronym 'XX]{Make sure to enter the correct
  conference title from your rights confirmation email}{June 03--05,
  2018}{Woodstock, NY}
\acmISBN{978-1-4503-XXXX-X/2018/06}

\usepackage{multirow}
\usepackage{multicol}
\begin{document}

\title{GCL-Sampler: Discovering Kernel Similarity for Sampled GPU Simulation via Graph Contrastive Learning}
%%  缩写：GCS；GraCS

\author{Jiaqi Wang}
\email{lingqi@mail.ustc.edu.cn}
\affiliation{%
  \institution{University of Science and Technology of China}
  \city{Hefei}
  \country{China}
}

\author{Jingwei Sun}
\affiliation{%
  \institution{University of Science and Technology of China}
  \city{Hefei}
  \country{China}}
\email{sunjw@ustc.edu.cn}

\author{Jiyu Luo}
\affiliation{%
  \institution{University of Science and Technology of China}
  \city{Hefei}
  \country{China}}
\email{luojiyu@mail.ustc.edu.cn}

\author{Han Li}
\affiliation{%
  \institution{University of Science and Technology of China}
  \city{Hefei}
  \country{China}}
\email{hanli06@mail.ustc.edu.cn}

\author{Guangzhong Sun}
\affiliation{%
  \institution{University of Science and Technology of China}
  \city{Hefei}
  \country{China}}
\email{gzsun@ustc.edu.cn}

\renewcommand{\shortauthors}{Wang et al.}

\begin{abstract}

GPU architectural simulation is orders of magnitude slower than native execution, necessitating workload sampling for practical speedups. Existing methods rely on hand-crafted features with limited expressiveness, yielding either aggressive sampling with high errors or conservative sampling with constrained speedups. To address these issues, we propose GCL-Sampler, a sampling framework that leverages Relational Graph Convolutional Networks with contrastive learning to automatically discover high-dimensional kernel similarities from trace graphs. By encoding instruction sequences and data dependencies into graph embeddings, GCL-Sampler captures rich structural and semantic properties of program execution, enabling both high fidelity and substantial speedup. Evaluations on extensive benchmarks show that GCL-Sampler achieves $258.94\times$ average speedup against full workload  with 0.37\% error, outperforming state-of-the-art methods, PKA ($129.23\times$, 20.90\%), Sieve ($94.90\times$, 4.10\%) and STEM+ROOT ($56.57\times$, 0.38\%).

\end{abstract}





\maketitle

\section{Introduction}

Graphics Processing Units (GPUs) are widely used in modern computing infrastructure. As GPU architectures grow increasingly complex, detailed performance modeling and design space exploration have become critical. GPU simulators, such as Accel-Sim\cite{accelsim}, GPGPU-Sim\cite{gpgpusim}, and HyFiSS\cite{hyfiss}, serve as indispensable tools for architects, enabling fine-grained analysis of microarchitectural behaviors, validation of novel hardware features, and evaluation of optimization strategies before silicon fabrication.

However, high-fidelity simulation inevitably suffers from severe performance penalties. Modern GPU simulators operate at speeds several orders of magnitude slower than native hardware execution. For large-scale workloads, particularly in machine learning where workloads may execute trillions of instructions, full simulation becomes prohibitively expensive, often requiring days or even weeks to complete a single run\cite{photon}. This simulation bottleneck fundamentally limits the scope and pace of GPU architecture
research, preventing architects from thoroughly exploring design alternatives or evaluating emerging workloads at scale.

\begin{figure}[ht]
  \centering
  \includegraphics[width=0.9\linewidth]{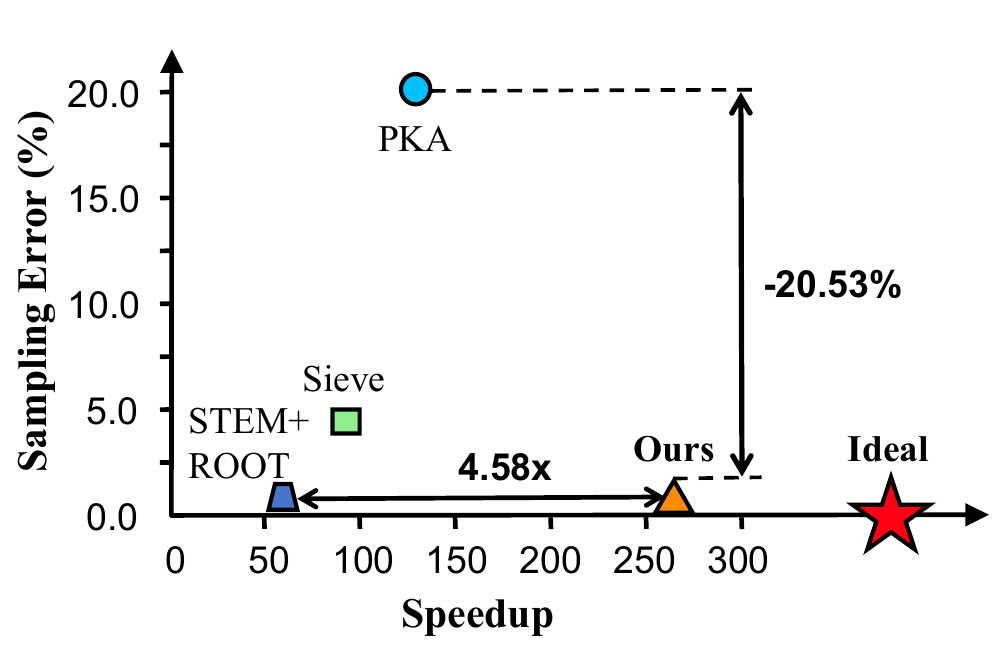}
  \caption{GCL-Sampler achieves near-ideal performance (red star) with minimal error and maximum speedup compared to existing methods.}
  \label{fig:intro}
\end{figure}

Workload sampling has emerged as a practical solution to accelerate GPU simulation by selecting a representative subset of execution intervals for detailed modeling while skipping the remainder. The fundamental challenge in sampling lies in identifying which execution intervals faithfully capture the overall behavior. If selected intervals fail to accurately represent the behavior of other intervals in their groups, simulation results will diverge from ground truth, leading to incorrect design decisions. If too many representative intervals are selected, the sampling approach fails to deliver meaningful speedups, undermining its practical value. Therefore, the central challenge in sampling methodology is to achieve high fidelity while maximizing simulation speedup.

Existing GPU sampling methods primarily rely on hand-crafted features to characterize kernel behavior and guide sample selection. Principal Kernel Analysis(PKA)\cite{pka} profiles twelve microarchitecture-independent execution characteristics like memory access patterns and instruction mixes to characterize kernels. These hand-crafted features are insufficient to adequately represent kernel behavior, leading to high sampling errors. Sieve \cite{sieve} addresses the accuracy issue of PKA through a conservative strategy. It first strictly partitions kernel invocations by their names, then stratifies each name group based on instruction count variability using the Coefficient of Variation (CoV). While this name-based constraint reduces errors compared to PKA, it prohibits clustering kernels with different names but similar performance, resulting in minimal speedup gains. Moreover, instruction count exhibits limited expressiveness and poor robustness, leaving Sieve vulnerable to high sampling errors. STEM+ROOT \cite{stem} similarly groups kernels by name and employs execution time distributions for fine-grained hierarchical clustering, but further reduces sampling errors by allowing multiple representative samples per cluster based on statistical error modeling. While this strategy improves robustness and reduces sampling errors, selecting more representatives lowers the overall speedup, making the approach less practical for scenarios requiring aggressive simulation acceleration.

These methods share a fundamental limitation: hand-crafted features lack sufficient expressiveness to capture the comprehensive behaviors of modern GPU programs. Due to their limited expressiveness, they face an inherent tradeoff: either accepting high sampling errors or resorting to overly conservative strategies that sacrifice achievable speedup to reduce errors.

Our key insight is that the structural and semantic properties of GPU kernel execution can be naturally encoded into Heterogeneous Relational Graphs (HRGs), where nodes and edges have different types to represent heterogeneous relationships in program structure. Compared to hand-crafted features, they can provide richer representation, thus enabling both low error and high speedup. Inspired by this, we propose GCL-Sampler, a GPU workload sampling framework that automatically learns high-quality kernel representations through Relational Graph Convolutional Networks (RGCN) \cite{RGCN} with contrastive learning \cite{CL}. We collect execution traces for GPU programs using NVBit \cite{nvbit}, construct per-kernel HRGs from these traces, and then train an RGCN to generate graph embeddings. These learned embeddings  reveal fine-grained behavioral similarities between kernels, enabling K-Means clustering \cite{kmeans} to select representative kernels for subsequent simulation. Figure ~\ref{fig:intro} illustrates that GCL-Sampler achieves both lower sampling error and higher speedup compared to existing approaches. 

In summary, this paper makes the following contributions:
\begin{itemize}
    \item \textbf{Novel Graph-Based Representation:} We present a sampling framework of RGCN with contrastive learning to provide high-quality graph embeddings for GPU kernels.

    \item \textbf{High-Fidelity, High-Speed Sampling:} GCL-Sampler achieves both state-of-the-art accuracy and speedup. Evaluated on 7,746 kernels across extensive workloads, GCL-Sampler achieves $258.94\times$ average speedup against full workload  with 0.37\% error, outperforming state-of-the-art methods, PKA ($129.23\times$, 20.90\%), Sieve ($94.90\times$, 4.10\%) and STEM+ROOT ($56.57\times$, 0.38\%). 

    \item \textbf{Comprehensive Experimental Validation:} We validate GCL-Sampler through comprehensive experiments across diverse benchmarks, multiple microarchitectural metrics, and cross-architecture evaluations, with end-to-end simulator integration demonstrating practical deployment in real-world simulation workflows.
\end{itemize}

\section{Background}

\subsection{SASS Trace}

NVIDIA GPUs execute SASS (Streaming Assembler) instructions that directly reflect hardware operation. A trace is a sequential record of executed instructions captured during program runtime. We employ NVBit, a dynamic binary instrumentation framework for NVIDIA GPUs, to collect SASS traces. SASS traces capture the true execution behaviors of GPU programs, including operations from closed-source libraries such as cuDNN \cite{cudnn} and cuBLAS \cite{cublas} that are otherwise opaque at higher abstraction levels.

\subsection{RGCN and Contrastive Learning}

We transform SASS traces into heterogeneous graphs with diverse node and edge types, making RGCN particularly suitable for modeling this representation. RGCN extends standard graph neural networks to handle heterogeneous graphs. In an RGCN, each edge type $r \in \mathcal{R}$ has associated learnable parameters, enabling the model to distinguish different relationships between nodes. The message aggregation at layer $k$ for node $v$ is computed as:
\begin{equation}
    h_v^{(k)} = \sigma\left(\sum_{r \in \mathcal{R}} \sum_{u \in \mathcal{N}_r(v)} \frac{1}{|\mathcal{N}_r(v)|} W_r^{(k)} h_u^{(k-1)} + W_0^{(k)} h_v^{(k-1)}\right).
\end{equation}
$\mathcal{N}_r(v)$ denotes neighbors of $v$ connected via edge type $r$, $W_r^{(k)}$ are relation-specific weight matrices, and $\sigma$ is an activation function. 

Contrastive Learning is a self-supervised learning paradigm that learns representations by contrasting positive pairs against negative pairs. In this paper, two augmented views from the same kernel form a positive pair, while views from different kernels constitute negative pairs. Given an anchor sample, contrastive learning pulls positive pairs closer while pushing negative pairs apart. This is achieved by maximizing agreement between positive pairs. 

\subsection{GPU Simulation}

GPU simulators create virtual hardware models at the software level to enable architecture exploration without physical prototyping. They are categorized into cycle-accurate and cycle-approximate approaches. Cycle-accurate simulators like GPGPU-Sim\cite{gpgpusim} and Accel-Sim\cite{accelsim} provide detailed microarchitectural modeling with high fidelity but extremely low simulation speeds, requiring hours to days for ML workloads. Cycle-approximate simulators\cite{pptgpu}\cite{hyfiss}\cite{ml1}\cite{ml2} trade accuracy for speed through higher-level abstractions but may lack robustness in representing complex hardware states. Even state-of-the-art simulators\cite{hyfiss} that balance accuracy and performance still face prohibitive simulation time for large-scale workloads, motivating the need for workload sampling techniques.

\section{The GCL-Sampler Methodology}

\begin{figure*}
  \includegraphics[width=\textwidth]{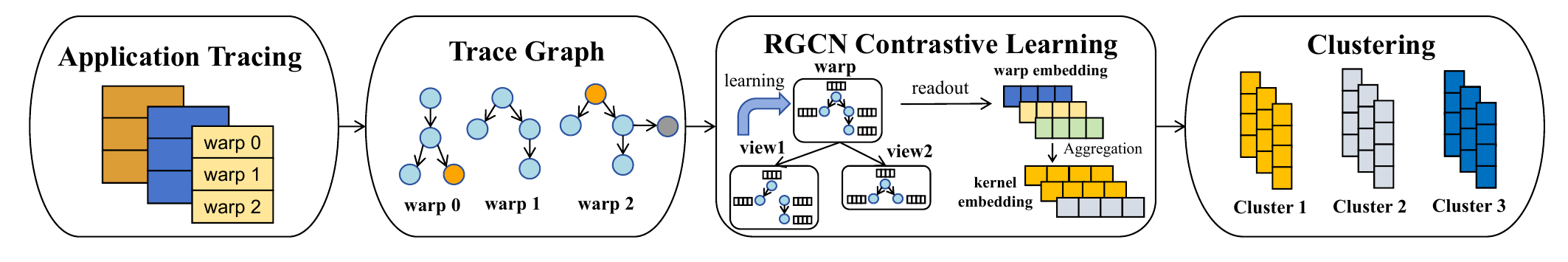}
  \caption{Overview of GCL-Sampler. GCL-Sampler transforms program traces into HRGs and leverages RGCN-based contrastive learning to generate kernel embeddings, followed by K-Means clustering.}
  \label{fig:overview}
\end{figure*}

In this section, we propose GCL-Sampler, a high-fidelity GPU kernel sampling framework based on RGCN contrastive learning. The core insight of GCL-Sampler is to leverage learned, high-quality graph embeddings to supersede traditional hand-crafted features, thereby enabling automatic discovery of inter-kernel similarities. 

Figure~\ref{fig:overview} illustrates an overview of our methodology. Our approach operates in four stages. First, we develop a tracing tool built on top of NVBit to collect the runtime trace of GPU programs. Second, we transform linear execution traces into HRGs that preserve both structural and semantic information. Third, these graphs are fed into an unsupervised RGCN contrastive learning framework. Given a program as input, this framework automatically learns embeddings to differentiate the kernels within the program. Finally, we apply K-Means to cluster these kernels for subsequent sampled simulation. While the learning phase incurs upfront preprocessing cost, this represents a one-time investment that amortizes across multiple design space exploration iterations.

\subsection{Application Tracing}
\label{tracing}

Accurate characterization of GPU kernel behavior requires capturing structural properties and dynamic characteristics. We build our tracing infrastructure on top of NVBit, NVIDIA's official dynamic binary instrumentation tool, which enables transparent interception of SASS instructions without requiring source code access.

\textbf{Trace Format and Organization.}
Our tracing tool generates one trace file per kernel invocation, recording executed instructions in temporal order. Each trace entry captures comprehensive information for each instruction, as Table~\ref{tab:trace_format} shows. This comprehensive format captures both static semantics and dynamic states necessary for constructing behavior-preserving graph representations. Raw traces organize instructions chronologically as they execute across threads. In post-processing, instructions from the same warp are grouped together while preserving their temporal ordering \cite{accelsim}, facilitating subsequent graph construction. 

\begin{table}
\centering
\caption{Trace components for each instruction.}
\label{tab:trace_format}
  \begin{tabular}{||c|c||}
      \hline
      Component & Description \\
      \hline
      \hline
      \(tb_x\), \(tb_y\), \(tb_z\) & CTA coordinates in the grid  \\
      \hline
      warp\_id & Warp identifier within the CTA  \\
      \hline
      PC & Program counter  \\
      \hline
      mask & 32-bit active lane mask  \\
      \hline
      \#dests & Number of destination registers  \\
      \hline
      [dest\_regs] & Destination register identifiers  \\
      \hline
      opcode & Instruction operation code (e.g., LDG)  \\
      \hline
      \#srcs & Number of source operands  \\
      \hline
      [src\_regs] & Source register identifiers  \\
      \hline
      mem\_width & Memory access width in bytes  \\
      \hline
      [dynamic\_values] & Register operands and memory addresses \\
      \hline
  \end{tabular}
\end{table}

\textbf{Scoped Trace Collection Strategy.}
To balance tracing overhead with representativeness, we adopt a selective instrumentation strategy inspired by HyFiSS\cite{hyfiss}, which instruments only a single representative SM per kernel invocation. Within the selected SM, we record complete traces for all cooperative thread arrays (CTAs). This scope captures intra‑SM resource contention and minimize the accuracy loss with reduced overhead.
The one-time fixed cost from trace generation and post-processing can be amortized across multiple design iterations, making the approach practical for architecture research workflows.

\subsection{Graph Construction from Traces}
\label{sec:graph_construction}

\begin{figure}[h]
  \centering
  \includegraphics[width=\linewidth]{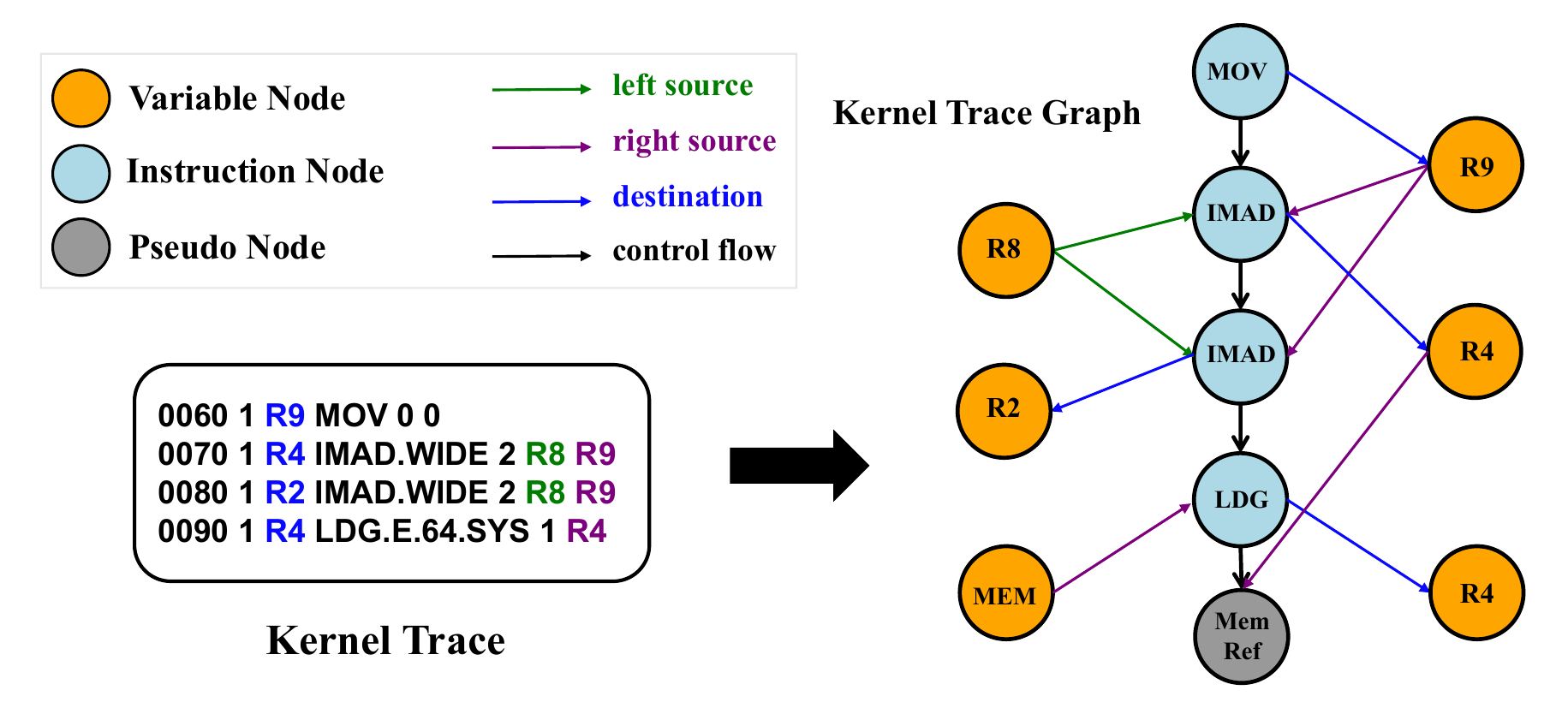}
  \caption{An example of graph construction from traces.}
  \label{fig:graph}
\end{figure}

Linear instruction traces lack explicit representation of structural relationships. To address this, we transform kernel traces into HRGs that encode control flow topology and data dependency. Figure ~\ref{fig:graph} illustrates an example of transforming a trace fragment into a graph. 

\textbf{Graph Definition.}
Formally, we represent each kernel $k$ as a directed heterogeneous graph $G_k = (V, E, \mathcal{R})$ constructed at warp granularity, where $V$ is the set of nodes, $E$ is the set of edges, and $\mathcal{R}$ defines typed edge relations. Each warp's execution trace is transformed into its own graph, and the final kernel graph is the union of all per-warp graphs. This warp-level construction naturally captures SIMT execution semantics while maintaining scalability.

\textbf{Graph Node.}
We define three categories of nodes to capture different aspects of execution: \textbf{(1)} \textbf{Instruction Node.} Each instruction node represents a SASS instruction(e.g., LDG). They are initialized with a token identifier(ID) corresponding to their opcode. \textbf{(2)} \textbf{Pseudo Node.} These nodes represent operations that occur within a single instruction but require explicit modeling to capture computational semantics(e.g., Mem Ref). They are also initialized with token IDs representing their operation type. \textbf{(3)} \textbf{Variable Node.} These nodes represent dynamic values during execution, such as register nodes and memory nodes. A new variable node is created only for a write, and all subsequent reads are connected to this most recent version, which is well exemplified by node \textit{R4} in Figure ~\ref{fig:graph}. If a variable node has no incoming edges, we initialize it with the value recorded in traces; otherwise, we initialize it to zero. RGCN will propagate values through data flow edges during forward passes, effectively "computing" the variable's value from its dependencies.

\textbf{Graph Edge.}
We define two edge categories to capture different dependency semantics: \textbf{(1) Control Flow Edges.}  They connect consecutively executed instructions in the trace, forming a directed path that reflects the actual runtime execution sequence. Since the trace captures the realized execution path, branch outcomes are implicitly encoded in this sequential structure, aligning with the hardware's SIMT execution model. \textbf{(2) Data Flow Edges.} These edges point from the source nodes to the result nodes, encompassing three types as shown in the legend of Figure ~\ref{fig:graph}.

\subsection{RGCN Contrastive Learning}
\label{sec:contrastive learning}

Having constructed the trace graphs, we now turn to the problem of generating high-quality embeddings for kernels. This section presents our unsupervised contrastive learning framework, which trains an RGCN to discover similarity patterns among kernels without relying on predefined labels or hand-crafted features. 

\subsubsection{Data Preprocessing}
\label{sec:preprocessing}

We perform two preprocessing steps for trace graphs: feature engineering to initialize node representations, and graph augmentation to generate contrastive views.

\textbf{Node Feature Engineering.}
Each node in trace graphs requires an initial feature vector. 
For instruction nodes, we construct a 64-dimensional vector by combining a dense embedding of its token ID with a positional encoding derived from its normalized PC value.
For variable nodes, we first generate a 32-dimensional embedding from their token ID. This static representation is then concatenated with an 8-dimensional statistical summary vector computed from its dynamic values, including mean, standard deviation, median, minimum, maximum, 25th percentile, 75th percentile, and skewness coefficient. The concatenation finally yields a 40-dimensional vector for each variable node.
Pseudo nodes are initialized with a 16-dimensional dense embedding of their token ID. Finally, all vectors are zero-padded to a uniform dimensionality of 64. 

\textbf{Graph Augmentation.}
Contrastive learning requires augmented views of each input graph where positive pairs remain semantically similar while negative pairs are distinguishable. We design an augmentation pool consisting of three strategies that maintain behavioral similarity between augmented views and original graphs. Node dropping randomly removes 15\% of nodes along with their incident edges. Edge dropping randomly eliminates 15\% of edges. Feature noise injection adds small random perturbations sampled from a 
Gaussian distribution($\sigma=0.01$) to node features. For each graph, we stochastically apply one or two augmentation strategies from this pool to produce two distinct contrastive views. These augmented graphs serve as input pairs for contrastive training. 

\subsubsection{RGCN Architecture}
\label{sec:encoder_arch}

The core of our embedding generation is an RGCN encoder, which is particularly suitable for handling heterogeneous graphs. It consists of three RGCN layers with an input dimension of 64, a hidden dimension of 128, and an output dimension of 256. Each RGCN layer applies basis decomposition to reduce the parameter count. The convolution output is sequentially processed by Layer Normalization, ReLU activation, and Dropout for regularization. The final RGCN layer omits dropout to preserve the full feature representation.
We apply mean pooling readout function after RGCN layers to aggregate nodes into warp embeddings, then average all warp embeddings to obtain the kernel embedding $\mathbf{z}_k \in \mathbb{R}^{256}$. 
During training, a two-layer multilayer perceptron(MLP) projection head maps $\mathbf{z}_k$ to 128 dimensions with ReLU and dropout, then to the final 64-dimensional output $\mathbf{z}'_k$.

\subsubsection{Training Loss}
\label{sec:loss_function}

We use the symmetric InfoNCE (Normalized Temperature-scaled Cross-Entropy) loss~ \cite{infonce} in training, which is widely adopted in contrastive learning, encouraging the model to pull together positive pairs while pushing apart negative pairs.

Given a batch of $B$ kernels, we generate two augmented views per kernel, producing embedding sets $\mathbf{Z}'_1$ and $\mathbf{Z}'_2$ of shape $[B, 64]$. After L2-normalization, we compute the cosine similarity matrix $\mathbf{S} \in \mathbb{R}^{B \times B}$ scaled by temperature $\tau$:
\begin{equation}
\mathbf{S} = \frac{1}{\tau} \mathbf{Z}'_1 (\mathbf{Z}'_2)^\top.
\end{equation}
The diagonal entries $\mathbf{S}_{ii}$ represent the similarity between positive pairs, while the off-diagonal entries $\mathbf{S}_{ij}$ ($i \neq j$) represent similarities between negative pairs.
The InfoNCE loss for the forward direction is computed as the cross-entropy loss over the similarity matrix:
\begin{equation}
\mathcal{L}_{\text{CE}}(\mathbf{S}) = -\frac{1}{B} \sum_{i=1}^{B} \log \frac{\exp(\mathbf{S}_{ii})}{\sum_{j=1}^{B} \exp(\mathbf{S}_{ij})}.
\end{equation}
To avoid biasing the optimization toward one augmentation direction, we compute the loss in both directions and average them:
\begin{equation}
\mathcal{L} = \frac{1}{2} \left[ \mathcal{L}_{\text{CE}}(\mathbf{S}) + \mathcal{L}_{\text{CE}}(\mathbf{S}^\top) \right].
\end{equation}

\subsection{Embedding Generation and Clustering}
\label{sec:embedding_generation}

After the contrastive learning phase, our trained RGCN model serves as a feature extractor to generate high-quality kernel representations for sampling. For each kernel graph, we directly utilize the 256-dimensional graph-level embedding produced by the readout function as the kernel's behavioral signature.

Kernel embeddings are subsequently fed into the K-Means clustering algorithm to identify representative simulation points. We select $K$ by maximizing the silhouette coefficient, which measures the compactness and separation of clusters without requiring ground-truth labels. When multiple values of $K$ yield comparable silhouette scores, we favor the smaller $K$ to maximize sampling speedup. This is naturally compatible with our contrastive embedding space, where similar kernels are already drawn together and dissimilar ones pushed apart. Once the optimal $K$ is determined, we select the first kernel invocation in each cluster as the representative simulation point.

\section{Experiment Setup}

\textbf{Evaluation Platform.}
All programs were compiled with NVCC 12.0, and ground-truth metrics were collected via NVIDIA Nsight Compute 2023.2.0.0 \cite{ncu}. Experiments were conducted on three hardware platforms with different GPU microarchitectures. Table~\ref{tab:hardware} summarizes the configurations. Platform P1 uses an NVIDIA RTX 2080Ti for trace generation and primary evaluation. Platforms P2 and P3 employ RTX 3080Ti and RTX 4090, respectively, to validate the generalization capability of our clustering approach across architectures. All systems run Ubuntu 24.04 LTS.
\begin{table}[t]
\centering
\caption{Hardware configurations for experiments.}
\label{tab:hardware}
\begin{tabular}{llcc}
\toprule
& \textbf{CPU (Intel)} & \textbf{GPU} & \textbf{Architecture} \\
\midrule
P1 & E5-2680 v4 & RTX 2080Ti & Turing \\
P2 & i9-12900KF & RTX 3080Ti & Ampere \\
P3 & W5-3435X & RTX 4090 & Ada Lovelace \\
\bottomrule
\end{tabular}
\end{table}
To demonstrate end-to-end applicability, we integrated GCL-Sampler with HyFiSS\cite{hyfiss}, validating that our sampling methodology can effectively accelerate real-world simulation workflows.

\textbf{Workloads.}
We evaluate GCL-Sampler on 11 programs encompassing 7,746 kernels from diverse benchmark suites: PolyBench\cite{polybench}, Rodinia\cite{rodinia}, Tango\cite{tango}, and large language model (LLM) workloads including \textit{qwen1.5}\cite{qwen1.5}, \textit{phi-2}\cite{phi2}, \textit{pythia}\cite{pythia}. This diverse set spans applications from scientific computing to modern AI inference, providing comprehensive coverage of GPU kernel behaviors.

\textbf{Model Configuration.}
Our RGCN model consists of 3 relational graph convolutional layers with ReLU activation functions. The hidden dimension of RGCN layers is set to 128, and the graph readout layer produces 256-dimensional graph embeddings. A projection head with a 128-dimensional hidden layer further transforms these embeddings into final 64-dimensional representations for training. The model handles 4 edge types corresponding to different control-flow and data-flow relationships in trace graphs. Training uses the AdamW optimizer\cite{adamw} with an initial learning rate of 7e-4, decayed via cosine annealing scheduling\cite{cos}, and the temperature ${\tau}$ of 0.05. We partition kernels into 80\% training and 20\% validation sets. Training was performed on an NVIDIA A100 80GB GPU. Training time scales with program size: for large-scale programs (e.g., \textit{phi-2} from LLM benchmarks), approximately 12 minutes are required per 100 kernels, while smaller programs (e.g., \textit{lu} from PolyBench) require only 35 seconds per 100 kernels.

\textbf{Baseline Methods.} 
We compare GCL-Sampler against three state-of-the-art kernel-level sampling methods: PKA\cite{pka},  Sieve\cite{sieve} and STEM+ROOT\cite{stem}. PKA employs the same K-Means clustering strategy as our work. Sieve follows its original design, choosing the first kernel with the maximum CTA count within each cluster as the representative. For STEM+ROOT, we follow its original design and set the error bound $\epsilon$ to 0.25. We do not include Photon\cite{photon} since it is implemented on AMD GPUs.

\textbf{Evaluation Metrics.} 

(1) Accuracy. The accuracy of GCL-Sampler is measured by the absolute cycle count difference between the sampled cycle count and the total  cycle count, normalized to the total cycle count: 
\begin{equation}
    \text{error} = \frac{|C_{full} - C_{sampled}|}{C_{full}} \times 100\%.
\end{equation}
We apply the same methodology to compute the sampling error for other microarchitectural metrics, including L1 cache hit rate, L2 cache hit rate, achieved occupancy, and IPC.

(2) Speedup. We quantify speedup as the ratio of the total kernel execution time for the full execution divided by the total kernel execution time for all representative kernel invocations:
\begin{equation}
    \text{speedup} = \frac{\text{Execution\_Time}_{full}}{\text{Execution\_Time}_{sampled}}.
\end{equation}

(3) Cross-Architecture Robustness. We evaluate generalization across architectures by applying sampling decisions made on P1 to different GPU architectures (P2 and P3), measuring whether errors are consistently low across microarchitectural variations.

\section{Evaluation}

\subsection{Accuracy and Speedup}
\label{5.1}

\begin{figure}[h]
  \centering
  \includegraphics[width=\linewidth]{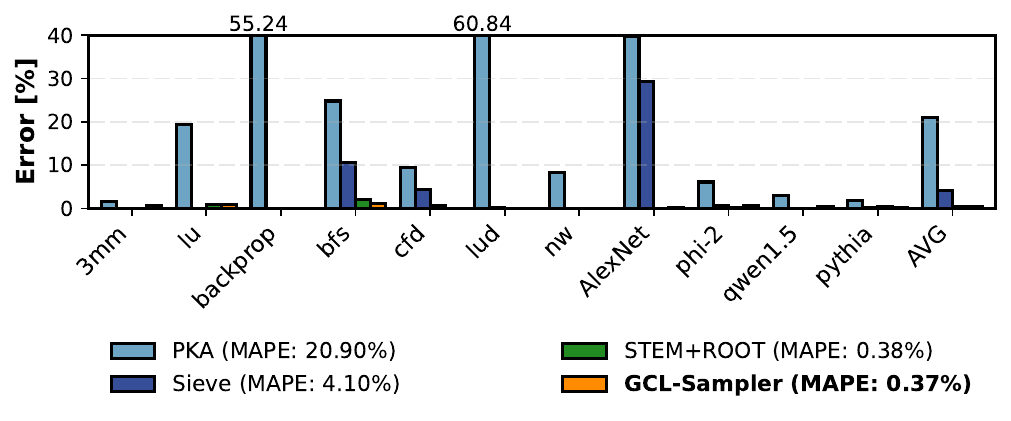}
  \caption{Sampling error of GCL-Sampler, PKA, Sieve and STEM+ROOT.}
  \Description{A woman and a girl in white dresses sit in an open car.}
  \label{fig:accuracy}
\end{figure}

\begin{figure}[h]
  \centering
  \includegraphics[width=\linewidth]{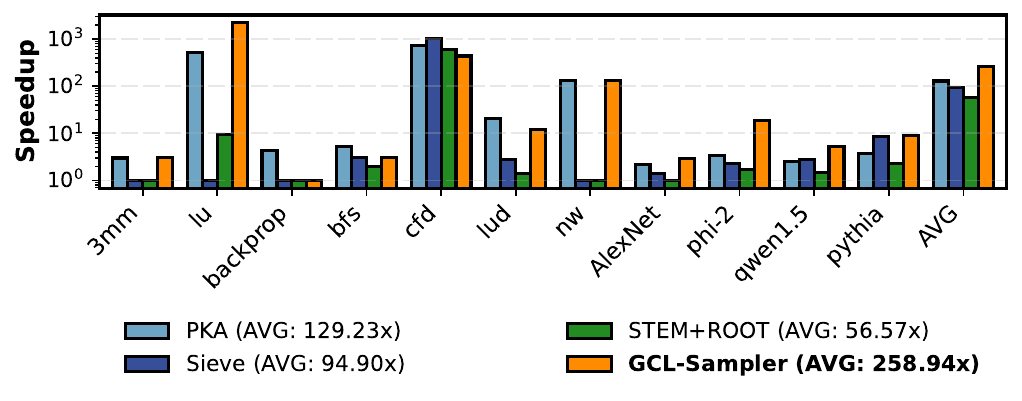}
  \caption{Speedup of GCL-Sampler, PKA, Sieve and STEM+ROOT.}
  \Description{A woman and a girl in white dresses sit in an open car.}
  \label{fig:speedup}
\end{figure}

Figure ~\ref{fig:accuracy} and Figure ~\ref{fig:speedup} present the sampling error and speedup achieved by GCL-Sampler compared to PKA, Sieve and STEM+ROOT across representative workloads, respectively. Our approach achieves an average error of 0.37\% with 258.94$\times$ speedup,  outperforming PKA (20.90\% error, 129.23$\times$ speedup), Sieve (4.10\% error, 94.90$\times$ speedup) and STEM+ROOT (0.38\% error, $56.57\times$ speedup). These results demonstrate that GCL-Sampler enables both high-fidelity sampling and maximum acceleration.

The key advantage of our method lies in accurately identifying performance-equivalent kernels regardless of their kernel names. In the \textit{nw} with 255 kernel invocations, GCL-Sampler, Sieve, and STEM+ROOT all achieve zero error. However, only our method groups kernels into two meaningful clusters, yielding approximately $130\times$ speedup. Both Sieve and STEM+ROOT rely on kernel-name-based grouping, which fails to find any reduction opportunity when all 255 kernels bear distinct names, completely negating the purpose of sampling. PKA matches our speedup but incurs 8.4\% error due to imprecise clustering. Similar patterns emerge in \textit{3mm} and \textit{lu}, where both Sieve and STEM+ROOT fail to identify reduction opportunities due to their reliance on kernel-name-based grouping, while PKA's coarse-grained metrics introduce notable errors. GCL-Sampler consistently delivers near-zero error with substantial speedup across these workloads.

PKA consistently exhibits high errors across multiple workloads, such as 55.2\% in \textit{backprop} and 60.8\% in \textit{lud}, which stems from its limited feature set that cannot adequately capture performance differences between kernels. Despite achieving considerable speedup, these sampling results are unreliable for architecture evaluation. Sieve, while maintaining lower overall error through kernel-name-based grouping, occasionally suffers from high errors due to its reliance on instruction count as the primary discriminative feature. In \textit{AlexNet}, this metric is unable to distinguish kernels with similar code size but different performance, resulting in 29.2\% error. STEM+ROOT addresses this limitation by allowing multiple representative samples per cluster. While this design enables STEM+ROOT to achieve consistently low error across all evaluated workloads, selecting more representative points per cluster comes at a significant cost to speedup. GCL-Sampler attains stable and robust low-error
performance while preserving maximum speedup.

\subsection{Cross-Architecture Validation}

A critical challenge in GPU sampling is ensuring that clustering decisions generalize across different hardware generations. Only by validating this cross-architecture transferability can we ensure that kernels selected on one platform remain representative on others, which is essential for applying sampling-based simulation to future hardware that has not yet been realized. We apply the clustering results obtained from P1 to ground-truth metrics collected on P2 and P3. In Table ~\ref{tab:combined_results}, we report the sampling errors and speedups across three generations of NVIDIA GPUs. Our method achieves consistently low average errors of 0.37\%, 1.50\%, and 1.22\% on P1, P2, and P3, respectively, while delivering speedups exceeding 200×. For the majority of workloads, the overall low error rates and performance trends show strong consistency between Turing, Ampere, and Ada Lovelace architectures. This demonstrates that kernels selected by GCL-Sampler on Turing remain equally representative on newer architectures. 

An exception is \textit{phi-2}, where error reaches 10.44\% on P2 and 9.74\% on P3. This anomaly mirrors an issue documented in prior work, PKA \cite{pka}, where profiler instrumentation triggers cuDNN to select different propagation algorithms at runtime based on performance heuristics, causing the GPU to execute different work across profiling runs and introducing measurement inconsistencies. Despite this library-level quirk, a 10\% error remains acceptable for architecture design space exploration.

\begin{table}[h]
\centering
\caption{Sampling error and speedup across GPU microarchitecture. P1 denotes Turing,  P2 denotes Ampere, and P3 denotes Ada Lovelace. E(\%)=Error(in \%), SU=Speedup.}
\resizebox{0.99\linewidth}{!}{\begin{tabular}{|l|cc|cc|cc|}
\hline
\multirow{2}{*}{Program} & \multicolumn{2}{c|}{P1} & \multicolumn{2}{c|}{P2} & \multicolumn{2}{c|}{P3} \\
\cline{2-7}
& E (\%) & SU & E (\%) & SU & E (\%) & SU \\
\hline
\hline
bfs & 1.07 & 3.04$\times$ & <0.01 & 3.17$\times$ & 0.05 & 3.51$\times$ \\
lu & 0.92 & 2224.67$\times$ & 0.11 & 1644.30$\times$ & <0.01 & 1658.56$\times$ \\
phi-2 & 0.63 & 18.41$\times$ & 10.44 & 13.42$\times$ & 9.74 & 14.01$\times$ \\
qwen1.5 & 0.38 & 5.19$\times$ & 2.20 & 3.76$\times$ & 1.93 & 3.81$\times$ \\
3mm & 0.71 & 3.05$\times$ & 1.61 & 3.00$\times$ & 0.67 & 3.03$\times$ \\
AlexNet & 0.18 & 2.95$\times$ & 0.14 & 3.27$\times$ & <0.01 & 3.25$\times$ \\
cfd & <0.01 & 438.95$\times$ & <0.01 & 426.98$\times$ & <0.01 & 407.39$\times$ \\
lud & <0.01 & 11.99$\times$ & <0.01 & 11.69$\times$ & <0.01 & 11.66$\times$ \\
nw & <0.01 & 130.13$\times$ & <0.01 & 127.84$\times$ & <0.01 & 128.60$\times$ \\
pythia & 0.18 & 8.96$\times$ & 2.05 & 5.28$\times$ & 1.08 & 5.25$\times$ \\
backprop & <0.01 & 1.00$\times$ & <0.01 & 1.00$\times$ & <0.01 & 1.00$\times$ \\
\hline
AVG & 0.37 & 258.94$\times$ & 1.50 & 203.97$\times$ & 1.22 & 203.64$\times$ \\
\hline
\end{tabular}}
\label{tab:combined_results}
\end{table}

\subsection{Validation on Microarchitectural Metrics}

\begin{figure}[h]
  \centering
  \includegraphics[width=\linewidth]{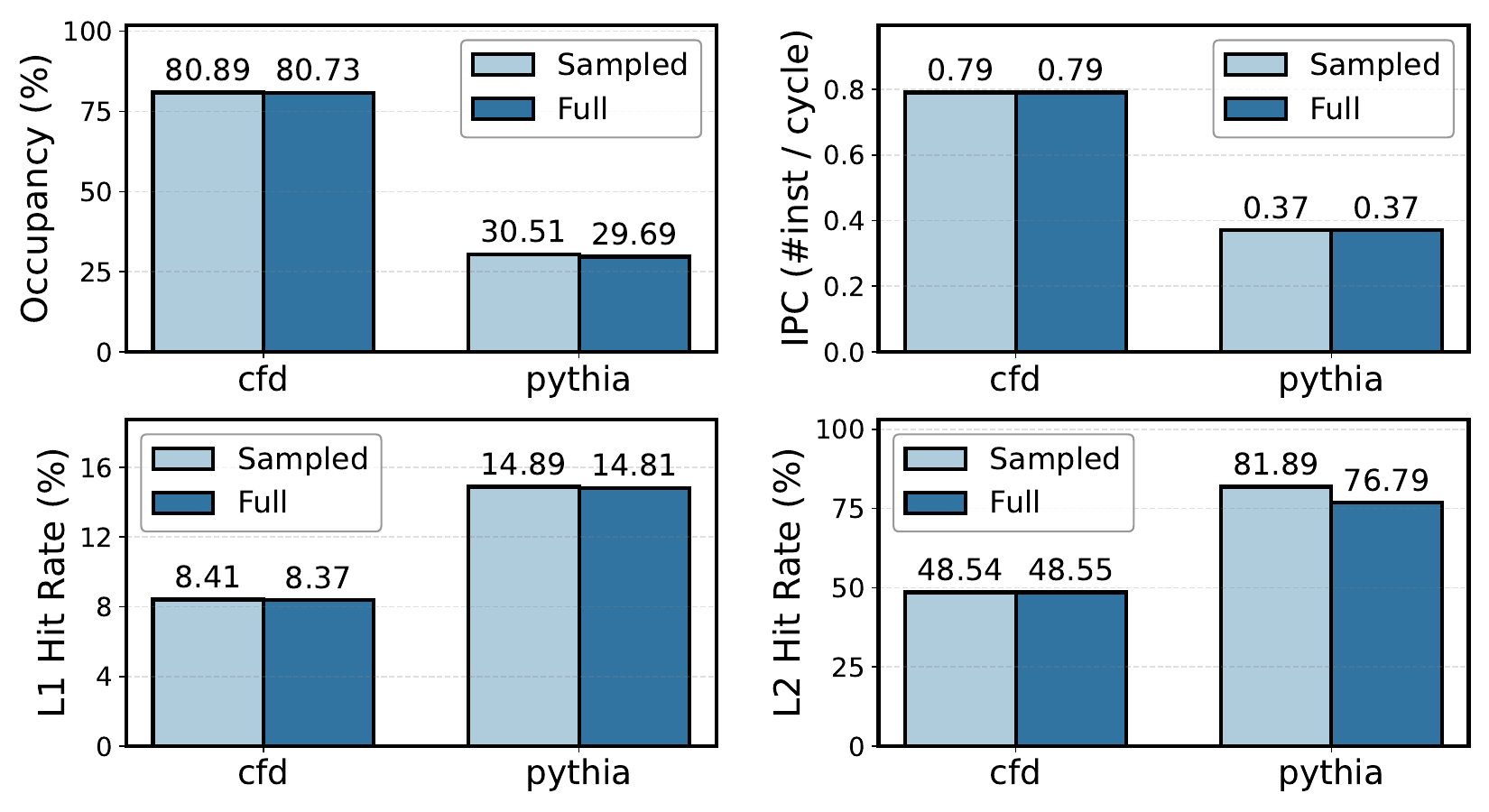}
  \caption{Microarchitectural metrics comparison between sampled and full simulation for \textit{cfd} and \textit{pythia} workloads.}
  \label{fig:metrics}
\end{figure}

Beyond execution cycles, we evaluate whether GCL-Sampler preserves accuracy across other microarchitectural metrics critical for understanding GPU behavior. Figure ~\ref{fig:metrics} compares achieved occupancy, IPC, L1, and L2 cache hit rates between full execution and sampled execution for \textit{cfd} from Rodinia and \textit{pythia} from our LLM suite. We calculate these metrics using a weighted sum over sampled kernels, applying the same approach used for GPU cycles in Section ~\ref{5.1}. The results show negligible deviation across all four metrics for both workloads. 
This accuracy demonstrates that our sampling approach captures diverse microarchitectural behaviors, not only for GPU cycles. These results confirm that architects can rely on GCL-Sampler for comprehensive performance analysis during early design stages, where understanding resource utilization and memory hierarchy behavior is essential for informed architectural decisions.

\subsection{End-to-End Simulation}

The core objective of sampling is to accelerate simulation by avoiding the prohibitive overhead of full workload execution. To validate practical speedup, we conduct an end-to-end simulation with HyFiSS, which accepts kernel-level requests compatible with our clustering. The integration process is straightforward. Given a workload with clustered kernels, we generate a simulation script that instructs HyFiSS to simulate only the representative kernel from each cluster, then reconstruct full workloads by scaling representatives by cluster weights.
The results confirm that GCL-Sampler achieves massive simulation speedup with negligible accuracy loss. For instance, simulating the full \textit{nw} workload in HyFiSS requires 22 minutes, while our sampled approach completes in just 10 seconds, delivering a 128× speedup with only 0.5\% cycle error. Large-scale LLM workloads such as \textit{phi-2} and \textit{ pythia} also achieve over 10× speedup. This validates that our clustering decisions translate effectively to actual simulator execution, enabling architects to rapidly explore design spaces that would otherwise be impractical due to simulation time constraints.
\section{Related Work}

Sampling techniques have long been essential for accelerating architecture simulation. Early CPU methods like SimPoint~ \cite{simpoint} used basic block vectors (BBVs) to identify representative execution intervals. LoopPoint~ \cite{looppoint} extended this via top-down program analysis, augmenting BBVs with runtime features capturing multi-threaded parallel behavior through loop structure analysis. More recently, NPS~\cite{nps} employed graph neural networks to learn execution embeddings from assembly code and runtime states, achieving improved accuracy over BBV-based approaches.

GPU programs require specialized sampling due to massive parallelism and heterogeneous execution. TBPoint~ \cite{tbpoint} introduced hierarchical sampling across inter-launch and intra-launch dimensions. Its inter-launch analysis is similar to the BBV-based clustering analysis process on CPUs, while intra-launch models thread block variations and warp state transitions through Markov processes. Yet, its reliance on full functional simulation for profiling limits scalability to large-scale workloads.
To address scalability challenges, Cesar A. Baddouh et al. proposed PKA, encompassing inter-kernel(PKS) and intra-kernel(PKP) sampling. PKS employs twelve architecture-independent profiling metrics to identify principal kernels, while PKP, which detects intra-kernel IPC stability, as the authors noted, provides limited acceleration.
Recognizing that PKA suffers from high intra-cluster variability and prohibitive profiling overhead, Sieve adopted a minimalist approach using only instruction count as a signature, with stratified sampling limited to identical kernel names. While achieving lower error than PKA, this constraint restricts grouping similar-performance kernels with different names, yielding minimal speedup, and the coarseness of instruction count as a feature leaves it vulnerable to high sampling errors when kernels share similar code size but exhibit distinct runtime behavior. STEM+ROOT addresses these shortcomings by replacing instruction count with execution time distribution as kernel signature, and employing statistical error modeling to adaptively determine the number of representative samples for each cluster based on its runtime variability. This achieves stable, low sampling error, though at the expense of significantly reduced speedup.
Photon, implemented on AMD GPUs, proposes online analysis for intra-kernel sampling without offline profiling but relies on GPU BBV for inter-kernel sampling, unable to interpret instruction semantics or capture dynamic values.

\section{Conclusion}

In this paper, we present GCL-Sampler, a GPU workload sampling framework that leverages RGCN-based contrastive learning to automatically discover kernel similarities from trace graphs. By encoding structural and semantic properties into learned embeddings rather than hand-crafted features, GCL-Sampler achieves both high fidelity and speedup. Extensive evaluations demonstrate that our approach reduces sampling error to 0.37\% while delivering 258.94$\times$ speedup, outperforming existing methods. Cross-architecture validation and end-to-end simulation confirm that GCL-Sampler provides a practical solution for accelerating GPU architectural simulation without sacrificing accuracy.

% \begin{acks}
% \end{acks}

\bibliographystyle{ACM-Reference-Format}
\bibliography{ref2026}

% \appendix








\end{document}